\documentstyle[12pt,blois]{article}

\begin{document}
\heading{\bf YOUNG DWARF GALAXIES AND COSMOLOGY}

\author{Trinh Xuan Thuan $^{1}$ and Yuri I. Izotov $^{2}$}
 {$^{1}$ University of Virginia, Charlottesville,
       USA.}  {$^{2}$ Main Astronomical Observatory, Kiev, Ukraine.}

\begin{bloisabstract}
We develop here the theme that extremely metal-deficient 
Blue Compact Dwarf Galaxies (BCDs), those with $Z$ $\leq$ $Z_\odot$/20, are 
young galaxies which did not start to form stars until $\sim$ 100 Myr ago.
They can be thus considered as primeval galaxies in our local volume of the 
universe, and are excellent laboratories for studying physical processes 
occuring at the time of galaxy formation in very metal-deficient environments.
We use BCDs to derive a new value of the primordial helium abundance 
$Y_p$ = 0.244$\pm$0.001, higher than previous determinations. This corresponds 
to a baryon mass fraction $\Omega_b$$h^2_{50}$ = 0.058$\pm$0.007. We discuss 
the problem of dust in a very low-metallicity environment and show 
how a large fraction ($\sim$ 75\%) of the young stars are not visible,
 and that any 
derivation of the cosmic star formation rate based only on optical/UV fluxes
would be an underestimate.  We also show 
how the loss of Ly$\alpha$ photons from starburst regions puts 
strong constraints on Ly$\alpha$ searches of high-redshift galaxies.

\end{bloisabstract}

\section{Introduction}

 The formation of galaxies is one of the most fundamental
 problems in astrophysics, and much effort has gone into the 
 search for primeval galaxies (PG). A possible definition of a primeval
  galaxy is a young system undergoing
its first major burst of star formation. It is now widely 
believed that the vast majority of galaxies underwent such a phase at 
redshifts $\sim$ 2 or greater. In most galaxy formation scenarios, 
young galaxies are predicted to show strong Ly$\alpha$ emission, 
associated with the cooling of the primordial gas and the subsequent 
formation of a large number of massive  ionizing stars
(Partridge \& Peebles 1967; Charlot \& Fall 1993).
 Yet, despite intensive searches, the predicted 
widespread population of Ly$\alpha$ primeval galaxies has remained 
elusive (Pritchett 1994). 

      Several objects have been put forward as possible PG candidates,
 ranging from high-redshift radio galaxies to Ly$\alpha$ 
emitters found around quasars and damped Ly$\alpha$ systems, 
mainly on the basis of very high luminosity and
star formation activity. However, most of these 
candidate PGs already contain a substantial amount of heavy 
elements, as evidenced by
the presence of strong P Cygni profiles and interstellar absorption in their
spectra (Steidel et al. 1996; Yee et al. 1996).  These
spectra are very similar to those of nearby starburst galaxies known to contain
old stellar populations (Leitherer et al. 1996).
Thus high-redshift galaxies discovered thus far  are not truly primeval.
Moreover, even if true PGs are discovered at high-redshift,
 it is difficult to study them 
in detail because of their extreme
faintness and very compact angular size. We propose here to take a different
approach to the PG problem. Instead of searching for very high-redshift 
galaxies in the process of forming, we look for nearby galaxies undergoing 
their first burst of star formation, and hence satisfying the above definition 
of a PG. The best candidates for such a search are blue compact dwarf 
galaxies (BCD).

      BCDs are  low-luminosity
 extragalactic objects with $M_B$ $\geq$ --18 where intense star
formation is presently occuring, as evidenced by their blue $UBV$ colors, and
their optical spectra which show strong narrow emission lines superposed on a
stellar continuum which is rising toward the blue, similar to spectra of HII
regions. Star formation in BCDs cannot be continuous but must proceed by 
bursts because of several observational constraints: 1) Gas is transformed 
into stars at the rate of approximately 1 $M_\odot$ yr$^{-1}$, so that the 
current burst cannot last more than about 10$^8$ yr before depleting the 
neutral gas supply of $\sim$ 10$^8$ $M_\odot$;
2) Optical-infrared colors of BCDs give burst ages of about 10$^7$ yr;
and 3) Population synthesis of UV spectra of BCDs give invariably jumps 
in the stellar luminosity function, indicative of starbursts (see Thuan 1991 
for a review).

      Ever since their discovery, the question has arisen whether BCDs are 
truly young systems where star formation is occuring for the first time, or old
galaxies with an old underlying stellar population on which the current 
starburst is superposed (Searle, Sargent \& Bagnuolo 1973). Thuan (1983)
carried out a near-infrared $JHK$ survey of BCDs and concluded that all the 
objects in his sample possessed an old underlying stellar population of K and 
M giants. That result was not unambiguous as the $JHK$ observations were 
centered on the star-forming regions and the near-infrared emission could be 
contaminated by light from young supergiant stars. The advent of CCD 
detectors allowed to look for the low-surface-brightness underlying component 
directly. Loose \& Thuan (1985) undertook a CCD imaging survey of a large BCD
sample and found that nearly all galaxies ($\geq$ 95\%) in their sample show an 
underlying extended low-surface-brightness component, on
which are superposed the high-surface-brightness star-forming regions.
Subsequent CCD surveys of BCDs have confirmed this initial result 
(Papaderos et al. 1996, Telles \& Terlevich 1997). 
Thus, most BCDs are not necessarily young galaxies. 
However, there was a hint that extremely metal-deficient BCDs 
do not contain an old stellar population and can be primordial. 
{\sl Hubble Space Telescope (HST)} 
imaging of I Zw 18, the most metal-deficient BCD known
 ($Z_\odot$/50, Searle \& Sargent 1972), to $V$ $\sim$ 26 by 
 Hunter \& Thronson (1995) suggests that the stellar population is 
dominated by young stars and that the colors of the underlying diffuse 
component are consistent with those from a sea of unresolved B or early
 A stars, with no evidence for stars older than $\sim$10$^7$yr.

      For more than 20 years, I Zw 18 stood in a class by itself. The BCD
metallicity distribution ranges from $\sim$ $Z_\odot$/3 to $\sim$ $Z_\odot$/50,
peaking at $\sim$ $Z_\odot$/10, and dropping off sharply for $Z$ 
$\leq$ $Z_\odot$/10. 
Intensive searches have been carried out to look for
low-metallicity BCDs but they have met until recently with limited success. 
 Several years ago,
a new BCD sample has been assembled by Izotov et al. (1993) from 
objective prism survey plates obtained with the 1m Schmidt telescope 
at the Byurakan Observatory of the Armenian Academy
of Sciences during the Second Byurakan Survey (SBS). The most 
interesting feature of the SBS is its metallicity
distribution (Izotov et al. 1992, Thuan et al. 1994): it contains significantly
 more low-metallicity BCDs than previous surveys. It has uncovered
about a dozen BCDs with $Z$ $\leq$ $Z_\odot$/15, more than doubling the 
number of such known low-metallicity BCDs and filling in the metallicity gap 
between I Zw 18 and previously known BCDs.

In section 2, we use the low-metallicity SBS sample together 
with other metal-deficient BCDs to study heavy element abundance ratios in 
very low-metallicity environments  
and to argue that all galaxies with a metallicity 
less than about 1/20 of solar metallicity are young, 
i.e. they did not start to form stars 
until about 100 Myr ago. In that sense very metal-deficient local 
BCDs are truly PGs,
and their study can shed light on galaxy formation at high redshift. 
Because metal-deficient BCDs are young, they constitute excellent laboratories 
for determining the primordial Helium abundance. We use a large sample of 
low-metallicity BCDs in section 3 to derive a new value 
of the primordial helium abundance which is appreciably higher than the
 previously 
accepted value, and a baryonic mass density more in agreement with 
other observations.  
We discuss next in detail two BCDs with $Z$ $\leq$ $Z_\odot$/20, 
SBS 0335--052 (section 4) and SBS 1415+437 (section 5). We show that 
photometric and spectroscopic data on these two galaxies
also point to a young age for the two BCDs, of less than 100 Myr. 
In the concluding sections we show how BCDs can shed light on 
cosmological issues such as dust in a very low-metallicity environment
(section 6), and the escape of Ly$\alpha$ photons from starburst galaxies  
(section 7).

\section{Galaxies with Z $\leq$ Z$_\odot$/20 are younger than 100 Myr}

\subsection{$\alpha$-elements}

The study of the variations of one chemical element relative 
to another is crucial for our understanding of the chemical evolution
of galaxies and for constraining models of stellar nucleosynthesis and
the shape of the initial mass function. In the case of BCDs,
it is  particularly important for understanding their  
evolutionary status, whether they are young or old.
Izotov \& Thuan (1999) have obtained very high-quality ground-based 
spectroscopic 
observations of 54 supergiant H II regions in 50 low-metallicity blue
compact galaxies with oxygen abundances 12 + log O/H between 7.1 and  8.3
($Z_\odot$/50 $\leq$ $Z$ $\leq$ $Z_\odot$/4).  They
use the data to determine abundances for the
elements N, O, Ne, S, Ar and Fe. They also analyze {\sl Hubble Space 
Telescope} ({\sl HST}) Faint Object Spectrograph archival spectra of 
10 supergiant H II regions to derive C and Si abundances in a subsample of 7  
BCDs. 
 The best studied and most easily observed element in BCDs is oxygen. 
Nucleosynthesis theory predicts it to be produced only by high-mass
($M$ $>$ 9 $M_\odot$) stars. We 
shall use it as the reference chemical element and consider the behavior of 
heavy element abundance ratios as a function of oxygen abundance.
Figures 1 d,e,f,g show the dependence of the 
abundance ratios Ne/O, Si/O, S/O and Ar/O on oxygen abundance.
The elements neon, silicon, sulfur and argon are all products of
$\alpha$-processes during both hydrostatic and explosive nucleosynthesis
in the same massive stars which make oxygen. Therefore, the Ne/O, Si/O,
S/O and Ar/O ratios should be constant and show no dependence on the oxygen 
abundance. As predicted by stellar nucleosynthesis theory,
 {\it none} of the above heavy element-to-oxygen abundance ratios
  depend on oxygen
abundance. The mean values of these element abundance ratios are directly
 related to the stellar yields and thus provide strong constraints on the 
theory of massive stellar nucleosynthesis
 (Thuan et al. 1995, Izotov \& Thuan 1999).

\def\baselinestretch{0.5}
\begin{figure}[t]
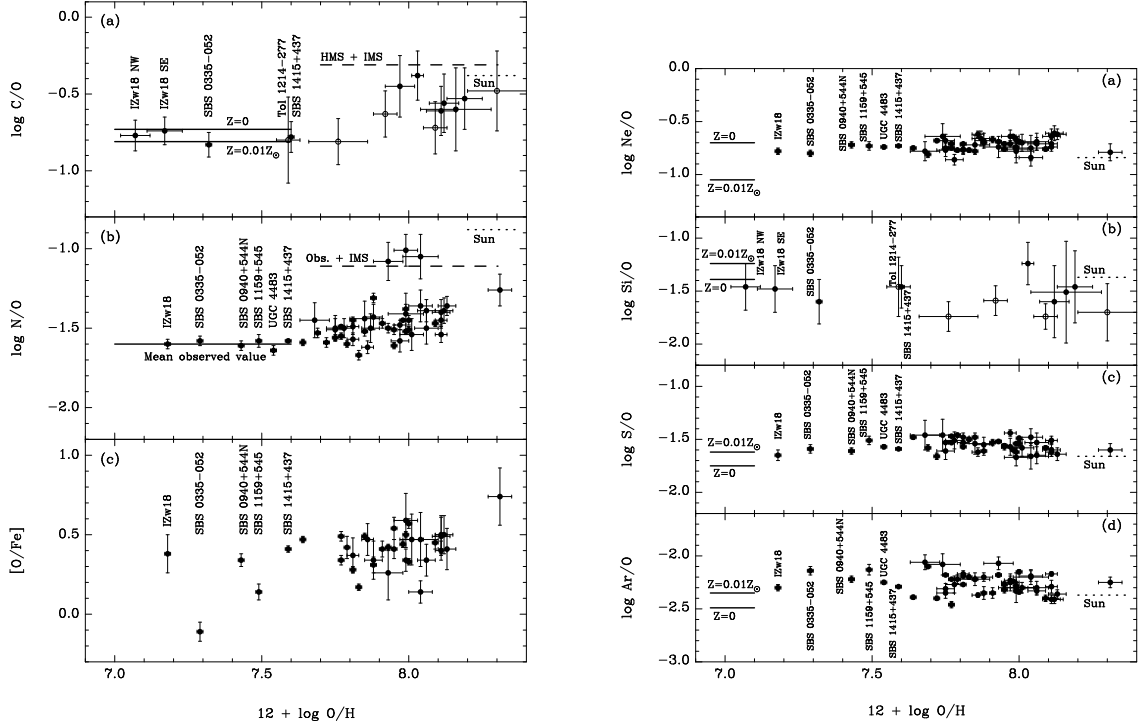

\centering
 \vbox{
	\includegraphics{thuan-fig1l.ps}}
	\includegraphics{thuan-fig1r.ps} 
\vspace{10.cm}
\caption[]{\label{Fig1}{\bf Left panel:} C/O, N/O and Fe/O abundance
ratios vs oxygen abundance for the sample of low-metallicity BCDs from
Izotov \& Thuan (1999). Solid lines for C/O are theoretical predictions
from high-mass stars (HMS) evolution (Woosley \& Weaver 1995); the dashed
line shows the  ratio predicted from the evolution of HMS and intermediate-mass
stars (IMS) (Renzini \& Voli 1981). The solid line for N/O is the mean 
observed ratio adopted as the primary nitrogen-to-oxygen abundance ratio
produced by massive stars.
{\bf Right panel:} Ne/O, Si/O, S/O and Ar/O abundance ratios vs oxygen
abundance for the same sample of BCDs. Solid lines are ratios predicted
from HMS evolution.}
\end{figure}
\def\baselinestretch{1.0}

\subsection{Iron, Carbon and Nitrogen}

  Because O, Ne, Si, S and Ar are made in the same high-mass stars, their 
abundance ratios with respect to O are constant and not sensitive to
the age of the galaxy. By contrast, C, N and Fe can be produced by both high 
and intermediate-mass (3 $M_\odot$ $\leq$ $M$ $\leq$ 9 $M_\odot$) stars,
and their abundance ratios with respect to O 
give important information on the evolutionary status of BCDs. 
The constancy of [O/Fe] for the BCDs  and its high
value compared to the Sun (Fig.1 c) suggests that all iron was produced by 
massive stars, i.e. in SNe II only. Since the time delay between 
iron production from SNe II and SNe Ia is about 1 -- 2 Gyr, it is likely 
that BCDs with oxygen
abundance less than 12 + log O/H $\sim$ 8.2 ($Z_\odot$/5)
 are younger than 1 -- 2 Gyr.
 
 The behavior of the C/O and N/O ratios as a function of oxygen 
abundance (Fig.1 a,b) puts more stringent constraints on the age of BCDs
(Thuan et al. 1995, Izotov \& Thuan 1999).
The behavior of the C/O and N/O ratios
is very different whether the BCG has 12 + log O/H smaller or greater than 7.6
($Z_\odot$/20).
The remarkably small scatter of the C/O and N/O abundance ratios in BCDs 
with $Z$ $\leq$ $Z_\odot$/20 rules out any time-delay model 
in which O is produced first by massive stars and C and N are produced later by 
intermediate-mass stars, and supports a common origin of C,
N and O in the same first-generation massive stars. Thus, it is very 
likely that the presently observed episode of star formation in BCDs with $Z$
 $\leq$ $Z_\odot$/20 is the first one in the history of the galaxy and the age of 
the oldest stars in it do not exceed $\sim$ 40 Myr, the lifetime of a 
9 $M_\odot$ star. 
    The conclusion that BCGs with $Z$ $\leq$ $Z_\odot$/20 are young is 
supported by the analysis of {\sl HST} WFPC2 images of some of these galaxies:
I Zw 18 ($Z_\odot$/50, Hunter \& Thronson 1995), SBS 0335--052 ($Z_\odot$/41,
 Thuan et al. 1997, see section 4), 
SBS 1415+437 ($Z_\odot$/21, Thuan et al. 1999, see section 5), 
T1214--277 and Tol 65 (respectively $Z_\odot$/21 and $Z_\odot$/22, 
Izotov et al. 1999).
    
   The situation changes for BCDs with $Z$ $>$ $Z_\odot$/20. The scatter of the 
C/O and N/O ratios increases significantly at a given O abundance, which was 
interpreted by Thuan et al. (1995) and Izotov \& Thuan (1999) as due to the 
additional production of primary N by 
intermediate-mass stars, on top of the primary N production by high-mass stars. 
Thus, since it takes at least 500 Myr (the lifetime of a 2 -- 3 $M_\odot$ star)
for C and N to be produced by 
intermediate-mass stars, BCDs with $Z$ $>$ $Z_\odot$/20 must have had several 
episodes of star formation before the present one and they must be at least 
older than $\sim$ 100 Myr. This conclusion is in agreement with photometric 
studies of these higher metallicity BCDs which, unlike their very 
low-metallicity counterparts, have a red old instead of a blue young underlying 
stellar component (Loose \& Thuan 1985, Papaderos et al. 1996).    

\section{The primordial helium abundance}

We use the data of 
Izotov et al. (1994, 1997a) and Izotov \& Thuan (1998b) to construct a 
sample of 45 HII regions appropriate for the determination of  $Y_p$.
Our sample constitutes one of the largest and most homogeneous
(obtained, reduced and analyzed in the same way) data set now available for the 
determination of  $Y_p$. This is the same data as used to analyze 
heavy element abundances in section 2. 
Linear regressions of the $Y$ -- O/H and $Y$ -- N/H relations, with
$Y$s determined from a self-consistent treatment of the five brightest optical
He I emission lines, gives $Y_p$ = 0.244$\pm$0.001 
(Figures 2 a,b, Izotov \& Thuan 1998b).
This value agrees very well with 
the mean $Y$ of the two most metal-deficient BCDs
known [I Zw 18 ($Y_p$ = 0.242$\pm$0.009, Izotov \& Thuan 1998a) 
and SBS 0335--052 ($Y_p$ = 0.249$\pm$0.004, Izotov \& Thuan 1998b)], which 
is $\bar{Y}$ = 0.245$\pm$0.006. Values as low as 
$Y_p$ = 0.234 or $Y_p$ = 0.230, as those obtained by Pagel et al. (1992) and
Olive et al. (1997) are excluded. Part of the difference comes from the 
fact that previous workers have neglected underlying He I stellar
absorption which would artificially lower $Y_p$ (Izotov \& Thuan 1988a). 

\def\baselinestretch{0.5}
\begin{figure*}[t]
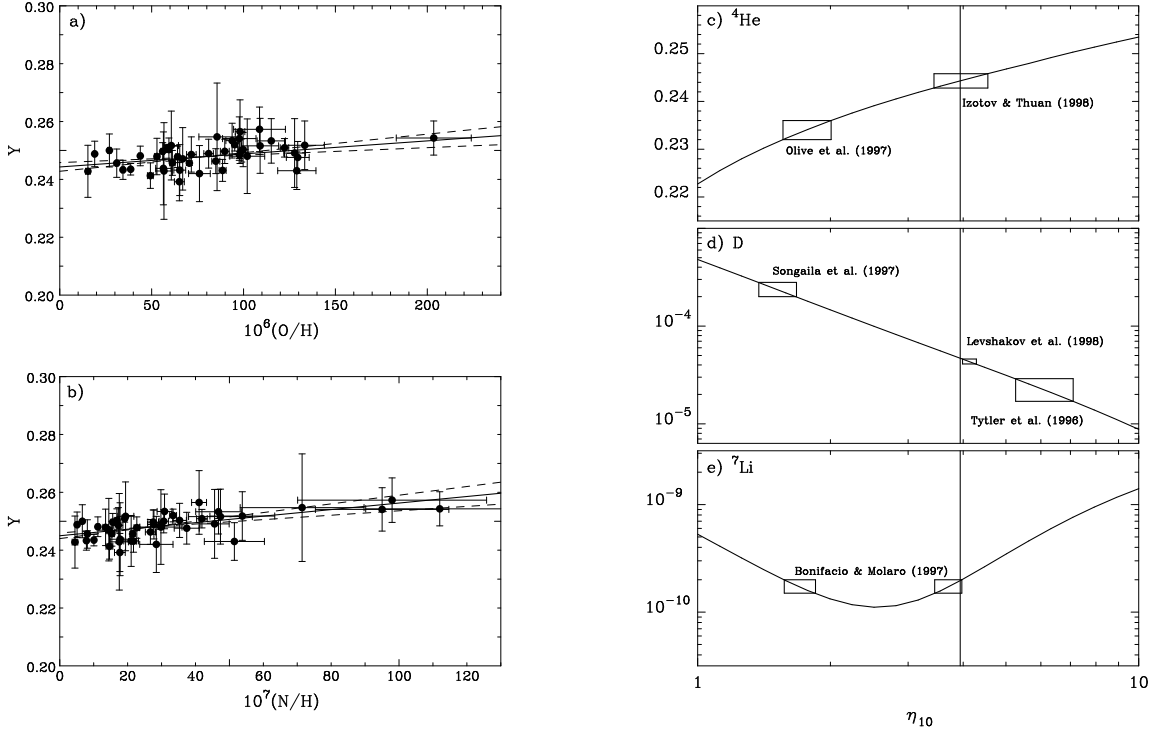

\centering
 \vbox{
	\includegraphics{thuan-fig2l.ps}}
	\includegraphics{thuan-fig2r.ps} 
\vspace{10.cm}
\caption[]{\label{Fig2}{\bf Left panel:}
Linear regressions of (a) the helium mass fraction
$Y$ vs. oxygen abundance O/H and (b) the helium mass fraction $Y$ vs.
nitrogen abundance for our sample of
45 H II regions (Izotov \& Thuan 1998b). The $Y$s are derived self-consistently 
by using the 5 brightest He I emission lines in the optical range.
Collisional and fluorescent enhancements, underlying He I stellar absorption and
Galactic Na I interstellar absorption are taken into account.
1$\sigma$ alternatives are shown by dashed lines.
{\bf Right panel:} 
The abundances of a) $^4$He; b) D and c) $^7$Li as a
function of $\eta_{10}$$\equiv$10$^{10}$$\eta$, where $\eta$ is the 
baryon-to-photon number ratio, as given by the standard
hot big bang nucleosynthesis model.}
\end{figure*}
\def\baselinestretch{1.0}

Our higher primordial helium mass fraction is consistent
with deuterium abundance measurements in high-redshift Ly$\alpha$ absorbing
systems by Tytler et al. (1996) as corrected by turbulence effects in the 
clouds by Levshakov et al. 1998), and lithium abundance measurements in 
low-metallicity halo stars (Bonifacio \& Molaro 1997). Figure 2c gives 
a baryon-to-photon ratio $\eta$= 4.0$\pm$0.5, which corresponds to 
a baryon mass fraction  $\Omega_b$$h^2_{50}$ = 0.058$\pm$0.007, in 
good agreement with the value derived from X-ray observations of clusters of 
galaxies (White \& Fabian 1995). 
We derive a slope $dY/dZ$ = 2.3$\pm$1.0, considerably smaller than those derived 
before.
 With this smaller slope and taking into account the errors,
 chemical evolution models with an outflow of well-mixed material can be
  built for star-forming dwarf
galaxies which satisfy all the observational constraints.  

\section{The young dwarf galaxy SBS 0335--052}

\subsection{Age of the stellar component}

This BCD with $Z$ = $Z_\odot$/41 is the second most metal-deficient galaxy known
(after I Zw 18).
HST WFPC2 images of the BCD show that most of the star formation in SBS 
0335--052 ($M_B$ = -- 16.7, $v$ = 4076 km s$^{-1}$)
occurs in 6 super-star clusters (SSCs) with --14.1 $\leq$ $M_V$ 
$\leq$ --11.9, within a region $\sim$ 520 pc in size (Figure 3, 
Thuan et al. 1997).
Later processing by Papaderos et al. (1998) of the same HST images reveals
several fainter clusters (not SSCs). The SSCs are roughly aligned in the SE-NW
direction, and there is a systematic reddening of the $V-I$ color  of the SSCs 
away from the brightest one, with a flattening of the color  of the clusters 
beyond 520 pc (Figure 4).

\def\baselinestretch{0.5}
\begin{figure*}[t]
 \vbox{
	\includegraphics{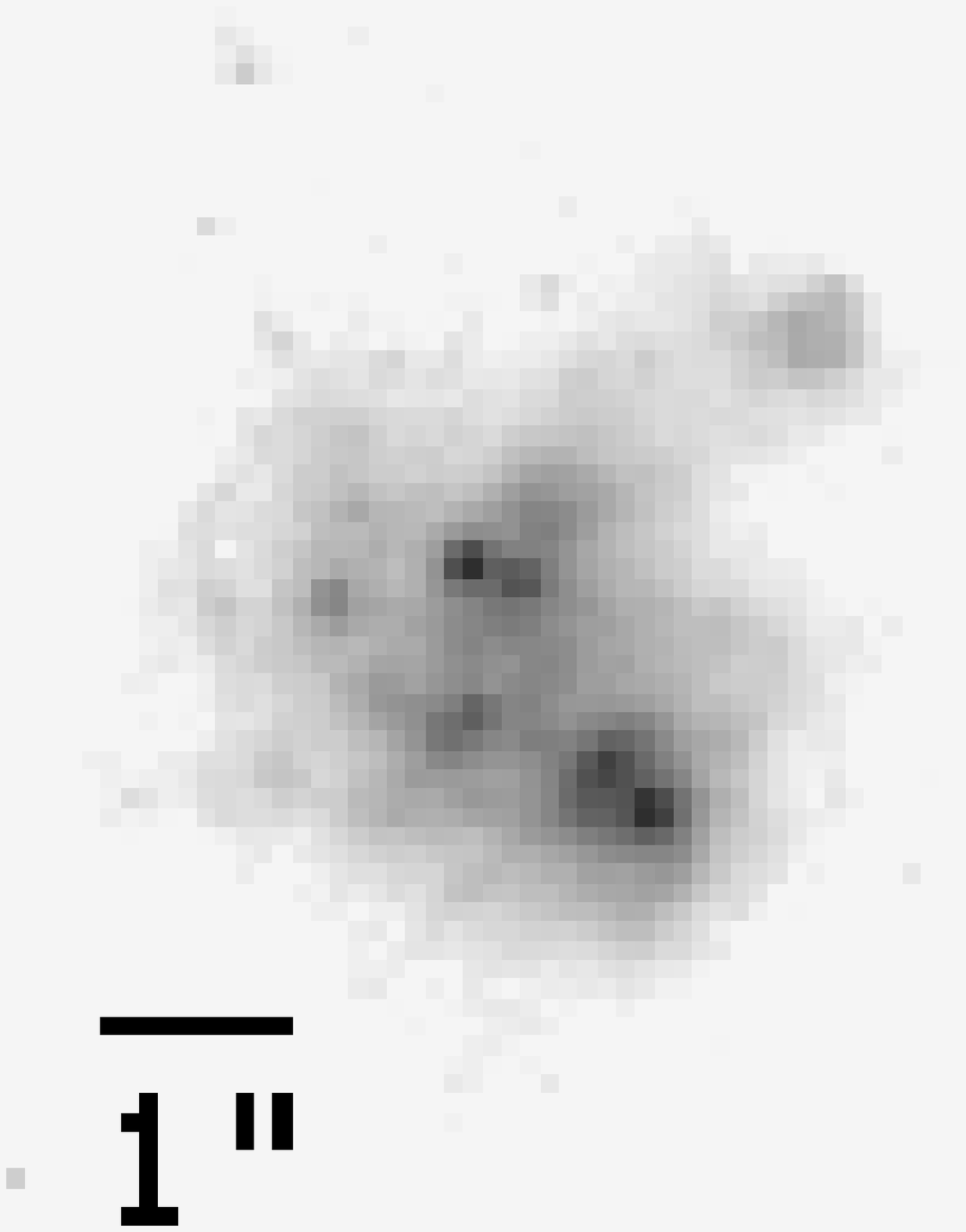}}
	\includegraphics{thuan-fig4.ps} 
\vspace{8.5cm}
{\label{Fig3_4}{\bf Fig. 3 (left):}
{\sl HST} WFPC2 $V$ image  of SBS 0335--052 showing the high surface brightness 
super-star clusters. At a distance of 54.3 Mpc, 1 arcsec corresponds to a linear size of 
263 pc.
{\bf Fig. 4 (right):} 
$(V-I)$ color vs. distance from the brightest super-star cluster at the SE tip
of SBS 0335--052.
The color gets redder with increasing distance.}
\end{figure*}
\def\baselinestretch{1.0}

\def\baselinestretch{0.5}
\begin{figure*}[t]
 \vbox{
	\includegraphics{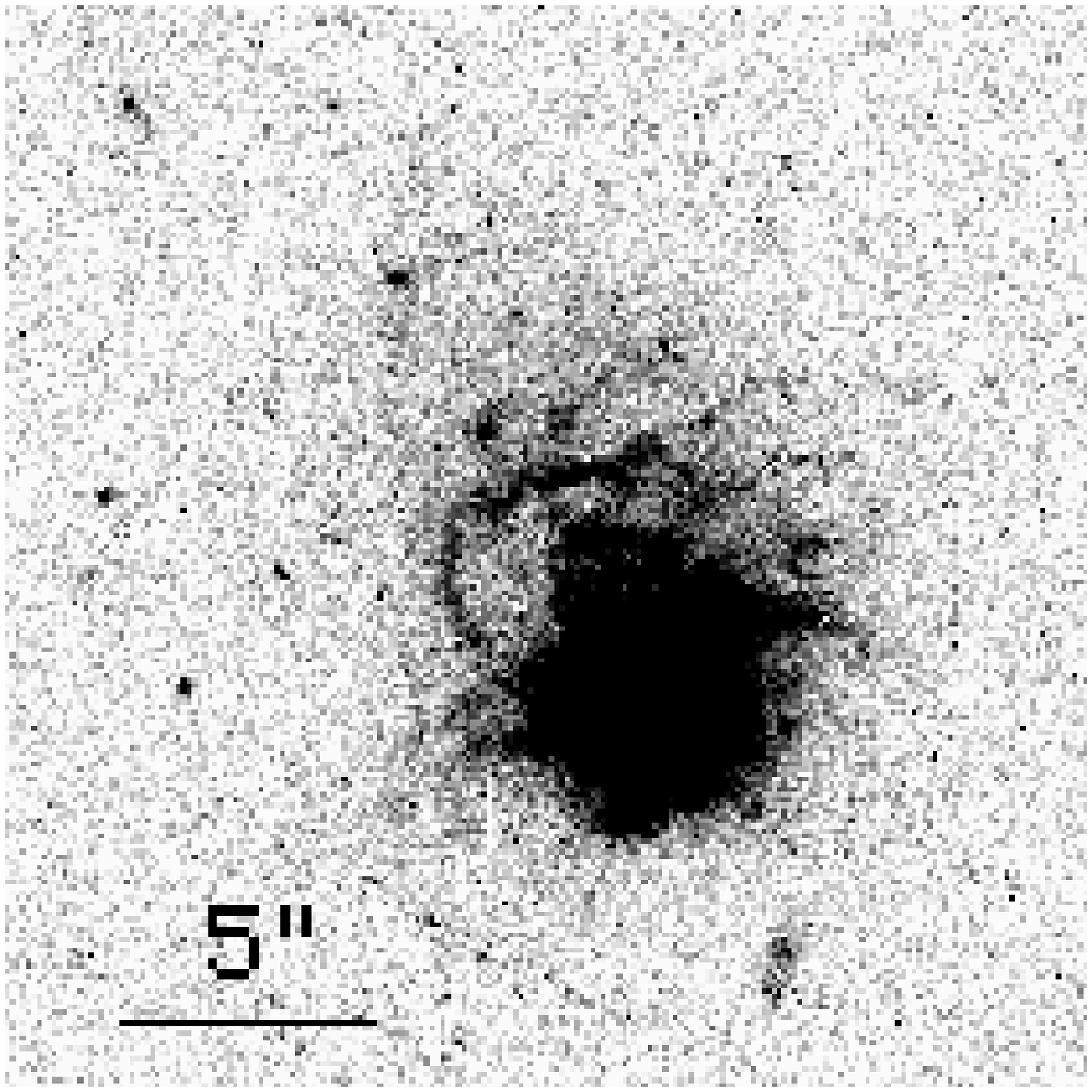}}
	\includegraphics{thuan-fig6.ps} 
\vspace{8.5cm}
{\label{Fig5_6}{\bf Fig.5 (left):}
Same $V$ image as in Fig.3, with the contrast adjusted to show the low surface 
brightness underlying component of SBS 0335--052.  The supershell delineating the 
large supernova cavity is clearly seen.

{\bf Fig.6 (right):} $(U-B)$ color profile (labeled E) of SBS 0335--052. The color is very blue and 
the profile flat.
 }
\end{figure*}
\def\baselinestretch{1.0}

Thuan et al. 
(1997) and Papaderos et al. (1998) attribute
most of the color variation of the SSCs
 to an age variation resulting from sequential propagating  star 
formation. The $V-I$ colors are consistent with the picture that star formation 
started at the location of the most distant cluster, 
some 1.8 kpc away from the 
location of the brightest and bluest cluster  at the South East 
end of the galaxy, at about 100 Myr ago and  propagated 
through the ISM to the latter, whose age is only $\sim$ 4 Myr, 
with an average speed of 
$\sim$ 18 km s$^{-1}$ (Thuan et al. 1997, Papaderos et al. 1998). Thus the 
star-forming clusters have ages between 4 and 100 Myr. Does SBS 0335--052 
possess an underlying older stellar population? 
The extended underlying component 
is shown in figure 5 where the contrast has been adjusted to display very 
low surface-brightness features. The unusually blue colors of this underlying
 component
(see the $U-B$ color profile labeled E in Figure 6) and its irregular, blotchy 
and 
filamentary structure suggest that a significant fraction of the light is of
 gaseous rather 
than stellar origin. A supershell of 380 pc radius can be seen delineating
 a large 
supernova cavity. However Izotov et al. (1997) found that the H$\beta$ 
equivalent 
width in the underlying component is $\sim$ 3 times lower than the value 
expected for pure gaseous emission, implying that two-thirds of the light 
comes from an underlying stellar population. Papaderos et al. (1998) 
have modeled the $UBVRI$ colors of this stellar component, after removal of 
the ionized gas contamination. They found that the colors are consistent 
with an underlying stellar population not older than 100 Myr.

\subsection {Age of the neutral component}

    Thus the stellar component in SBS 0355--052 is extremely young and 
the BCD is likely undergoing star formation for the first time. If this is the case,
the neutral gas envelope surrounding the BCD must also be very metal-deficient.
A 21 cm VLA map of the BCD (Pustilnik et al. 1999)  has shown it to be embedded 
in an extraordinarily large HI cloud seen nearly edge-on,
with dimensions some 64 by 24 kpc. This is 
to be compared with the typical size of HI envelopes around BCDs which is more 
like a few kiloparsecs in each dimension.  We can use the BCD as a background 
light source shining through the HI envelope to probe the physical conditions of the 
neutral gas. The Ly$\alpha$ line seen in absorption would give the column density 
of atomic hydrogen, while the OI $\lambda$1302 line would give the column density 
of the most abundant heavy element that remains neutral in the HI cloud. This 
would allow us to set limits on the O/H abundance ratio in the neutral gas.
 Figure 7 
shows the ultraviolet spectrum of SBS 0335--052 around the Ly$\alpha$ line obtained 
by Thuan \& Izotov (1997)
with the Goddard High Resolution Spectrograph aboard the {\sl Hubble Space Telescope (HST)}. 
A strong damped Ly$\alpha$ absorption line is seen along with several heavy element 
interstellar absoption lines such as OI $\lambda$1302, SiII $\lambda$1304 and 
SII $\lambda$1251, $\lambda$1254, and $\lambda$1259. The HI column 
derived by fitting the Ly$\alpha$ absorption profile is $N$(H I) =
(7.0$\pm$0.5)$\times$10$^{21}$ cm$^{-2}$, the highest
derived thus far for a BCD, and $\sim$2 times larger than in
I Zw 18 (Kunth et al. 1994).  Comparison
with high-resolution quasar spectra implies that the O I
$\lambda$1302 line along with other heavy element interstellar absorption lines
such as Si II $\lambda$1304 and S II $\lambda$1251, $\lambda$1254,
$\lambda$1259 are not saturated, which allow us to derive abundances.
Assuming that these lines originate in the H I gas, we derive 
extremely low abundances of
oxygen, silicon and sulfur, respectively 37000, 4000 and 116 times lower than
the solar values. The oxygen abundance is a whole 37 times lower than in the 
neutral gas of I Zw 18. However, these highly discrepant deficiency factors
between different elements suggest that the absorption lines are
produced, \underline{not in the H I, but in the H II gas}.
Adopting that hypothesis, the derived abundance from the $UV$ absorption lines
are then consistent with that derived from the optical emission lines ($Z$ 
$\sim$ $Z_\odot$/41). The conclusion that the heavy element absorption lines 
originate in the H II region is supported by the detection of several systems of
blueshifted S II $\lambda$1259, Si II $\lambda$1260, O I $\lambda$1302,
Si II $\lambda$1304, C II $\lambda$1335 absorption lines originating in 
fast-moving clouds with velocities up to $\sim$ 1500 km s$^{-1}$, and also by
the presence of heavy element 
absorption lines with excited lower levels. If this
conclusion holds, then the H I cloud in SBS 0335--052 is truly  
\underline{primordial}, unpolluted by heavy elements (Thuan \& Izotov 1997). 
In summary, all the known observational evidence suggests that SBS 0335--052 
is truly a young galaxy.

\section{Another young dwarf galaxy: SBS 1415+437}

\subsection {Age from color-magnitude diagrams} 

    Figure 8 shows  the {\sl HST} WFPC2 $I$ image of SBS 1415+437
   ($M_B$ = -- 14.0, $v$ = 607 km s$^{-1}$)  with the 
contrast level adjusted so as to show the
low surface-brightness underlying extended component
(Thuan et al. 1999). The galaxy has an  
elongated, comet-like shape with a bright H II region on its SW tip. Many
point sources  identified as luminous stars can be seen. To the SW
of the brightest H II region, two stellar clusters with resolved stars are
present. The luminous stars and the HII regions are aligned suggesting, 
just as in SBS 0335--052, propagating star formation (from the NE to the SW).
The mode of star formation in the two BCDs is different however. While 
SBS 0335--052 makes stars in luminous super-star clusters, star-formation in
SBS 1415+437  appears to be less extreme and is more similar to that 
in I Zw 18 (Hunter \& Thronson 1995). The superior spatial resolution of 
the {\sl HST} 
allows to resolve individual stars and construct color-magnitude diagrams to 
study the stellar populations in the BCD. To check the hypothesis of propagating 
star formation  , we have derived stellar ages for 6 separate regions 
in the BCD, labeled from I to VI as shown in Figure 8. 

  The $(V-I)$ vs. $I$ color-magnitude diagram (CMD) for each region 
  are shown in Figure 9 together
with stellar isochrones by Bertelli et al. (1994)
for a heavy element abundance equal to 1/20 the solar value.
 Each isochrone is marked
by the logarithm of the age in years. The region of
the  asymptotic giant branch (AGB) stars is
shown by a dashed line while the observational limits 
are shown by dotted lines. 
We adopt  $V$ = 27.5 mag and  $I$ = 27 mag as completeness
limits. Although there is evidence for spatial variations of the 
internal dust extinction, as a first approximation, we correct the CMDs for 
regions III -- VI by a constant extinction as derived by the Balmer decrement.
The CMDs for regions I and II have been not corrected for extinction for
lack of information.
   
Inspection of Figure 9 shows that there is a clear age gradient from region I to 
region VI. Region I is the youngest (about 5 Myr), containing only main-sequence 
stars. Region II is more evolved. It contains red 
supergiants and has an age of $\geq$ 10 Myr . 
Region III includes the brightest H II region
and shows a mixture of stellar populations. 
The bulk of the stars in region III have ages ranging between $\leq$ 10 Myr
and 100 Myr.  The red stars with $(V-I)$  between 1.0 and 1.8 
are likely to be red supergiants rather than AGB stars.
The stellar populations in regions IV - VI are 
similar to those in region III except for the fact that very young
populations with age less 10 Myr are no more present. 
We emphasize that the properties of the stellar populations
 in SBS 1415+437 are quite different from
those in other nearby low-metallicity dwarf galaxies with 
HST color-magnitude diagrams. In those dwarfs, very red AGB
stars are present  indicating a larger age
(e.g. Dohm-Palmer et al. 1997; Schulte-Ladbeck et al. 1998).
Two general conclusions can be obtained from the above color-magnitude
analysis: 
1) star formation in SBS 1415+437 is propagating from the NE to the SW;
2) there is no evidence for stars older than $\sim$ 100 Myr in the BCD.

\def\baselinestretch{0.5}
\begin{figure*}[t]
 \vbox{
	\includegraphics{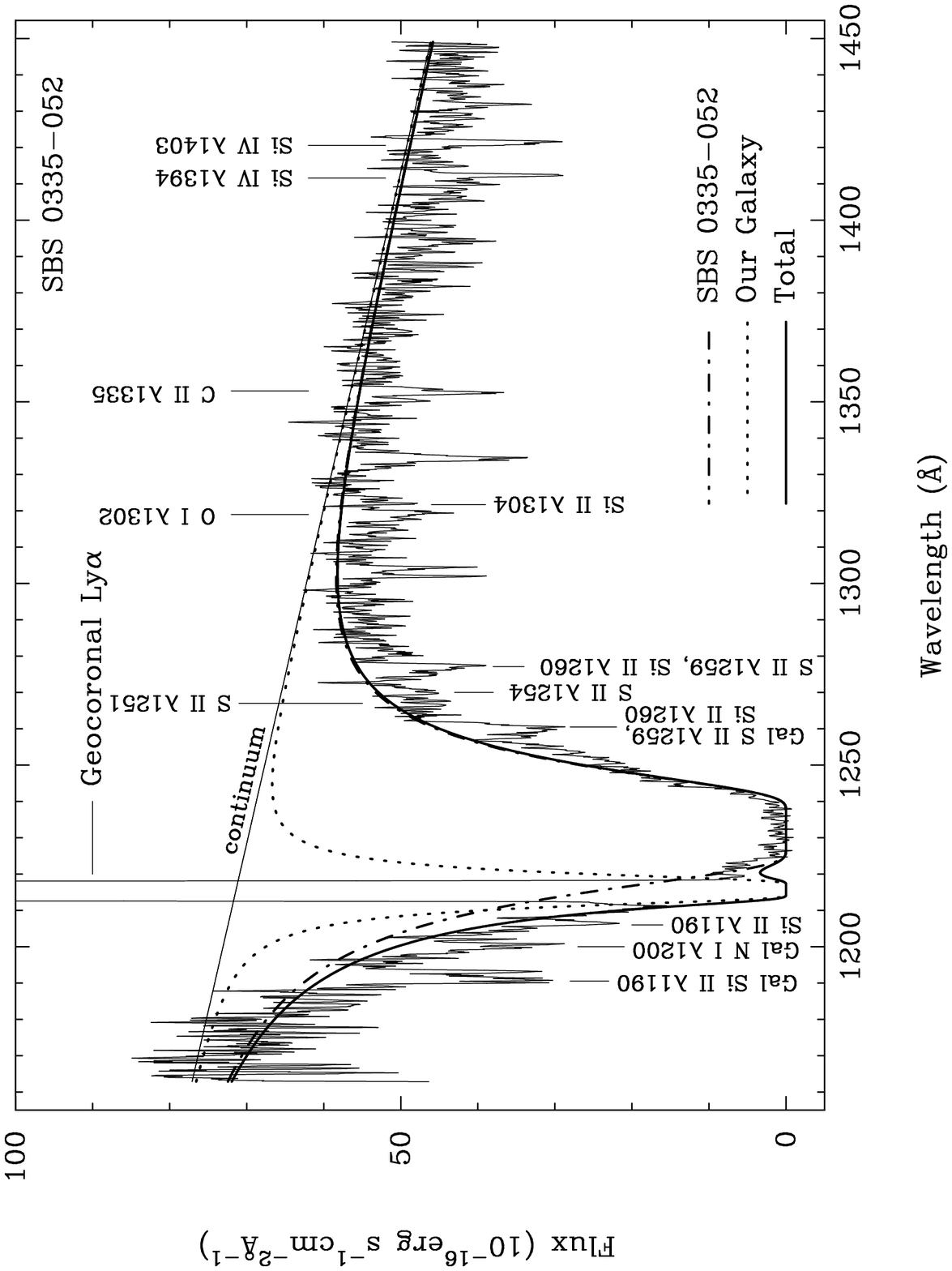}}
	\includegraphics{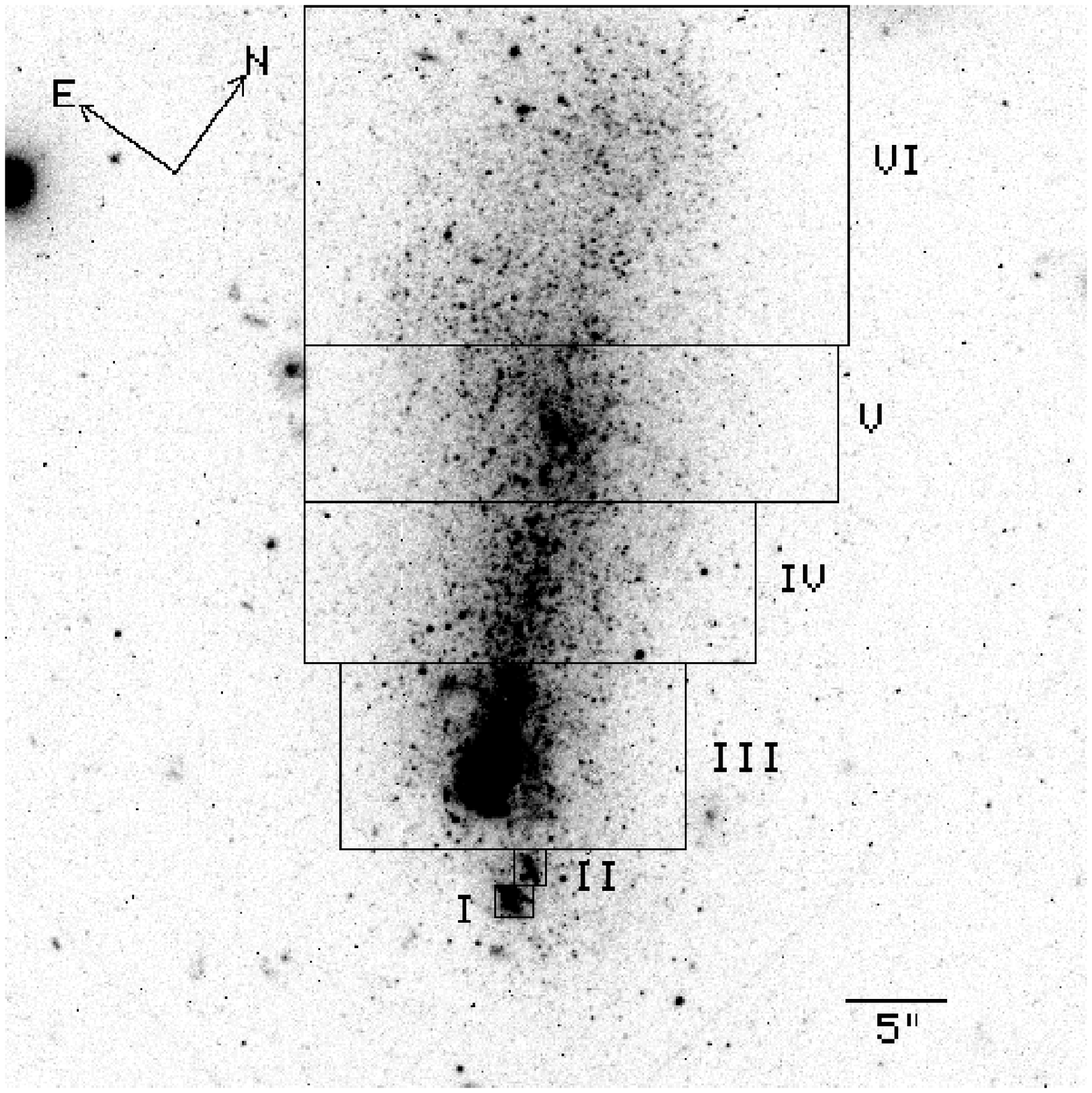}
\vspace{8.5cm}
{\label{Fig7_8}{\bf Fig.7 (left):}
HST GHRS spectrum of SBS 0335--052 around the Ly$\alpha$ line. A strong damped 
Ly$\alpha$ absorption is seen along with several heavy element interstellar absorption lines.

{\bf Fig.8 (right):} {\sl HST} WFPC2 $I$ image of SBS 1415+437. The shape is cometary-like.
A bright supergiant HII region at the SW end, resolved stars in the body of the galaxy 
and unresolved diffuse stellar emission can be seen. The different regions used for 
stellar population color-magnitude diagram analysis are marked.

 }
\end{figure*}
\def\baselinestretch{1.0}

\def\baselinestretch{0.5}
\begin{figure*}[t]
 \vbox{
	\includegraphics{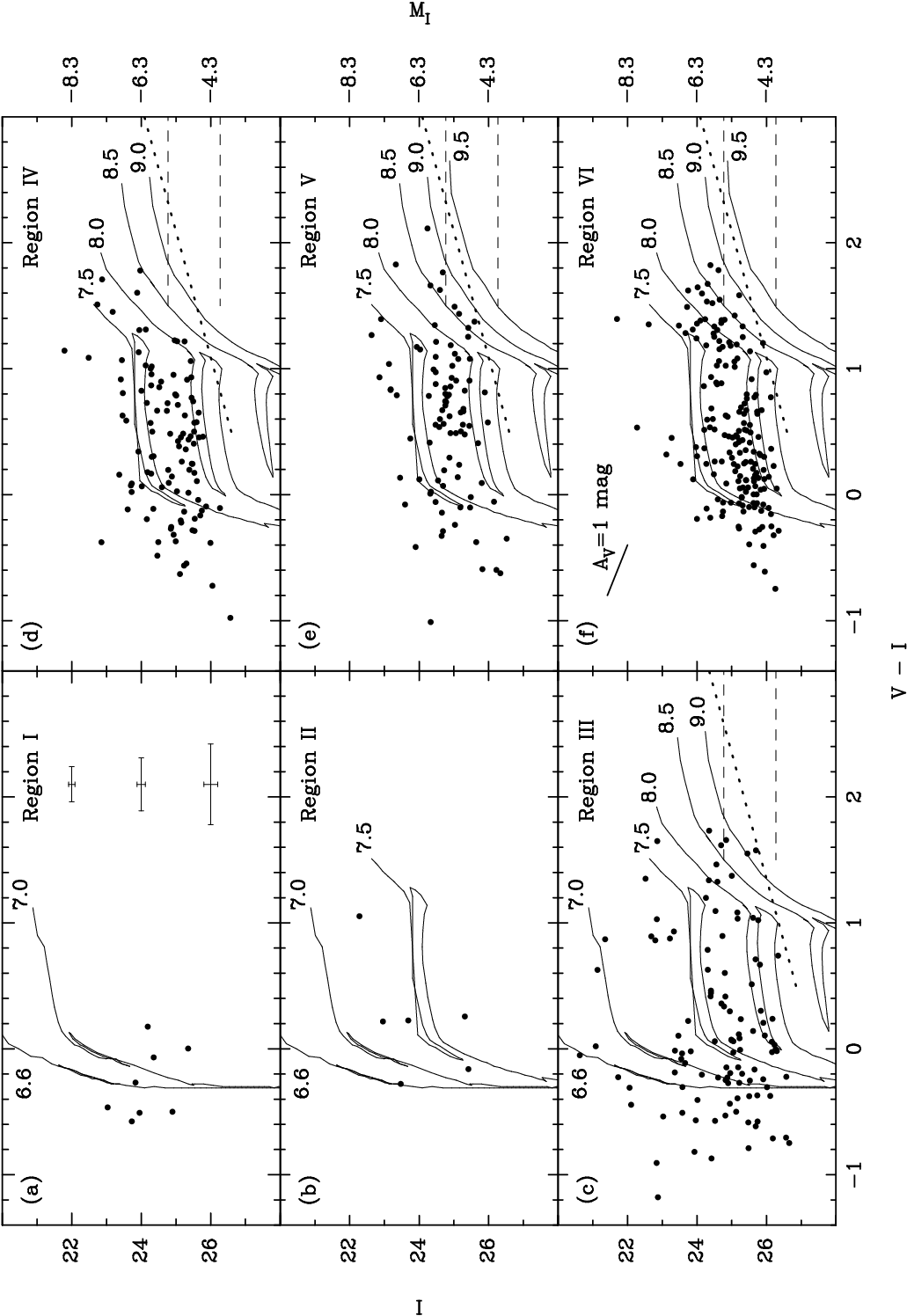}}
	\includegraphics{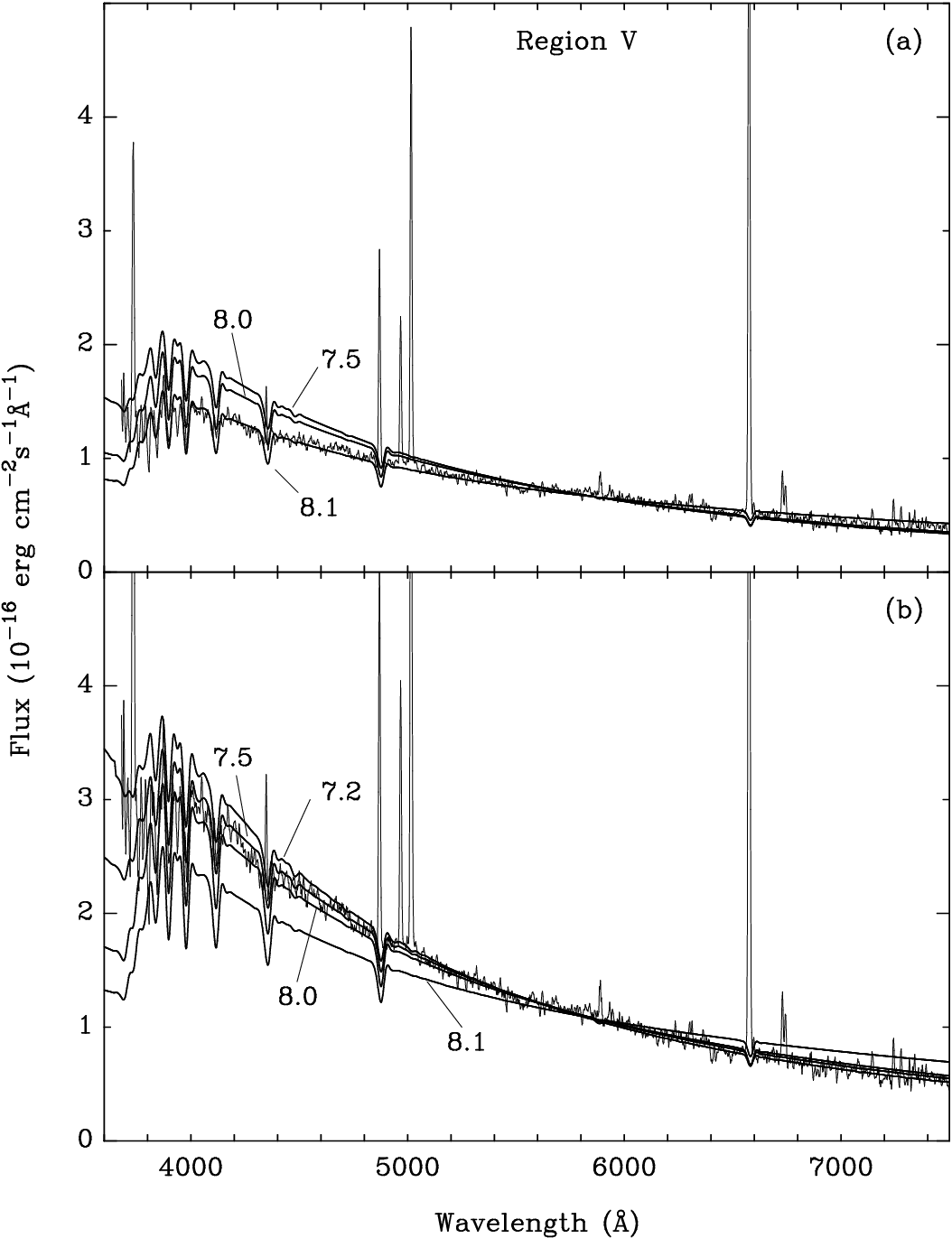}
\vspace{10.cm}
{\label{Fig9_10}{\bf Fig.9 (left):} Color-magnitude diagrams of stellar populations in the different 
regions of 
SBS 1415+437 as defined in Fig.8. There is a systematic age increase from region I 
to region VI, implying propagating star formation from the NE to the SW.
Typical photometric errors are shown in Fig. 9a.

{\bf Fig.10 (right):} MMT spectrum of region V (see Fig.7) of SBS 1415+437. The upper panel shows the 
spectrum uncorrected for extinction, while the lower panel shows the same spectrum 
corrected for extinction. The best-fit model gives an age between 30 and 100 Myr.
 }
\end{figure*}
\def\baselinestretch{1.0}

\subsection{Age from spectral evolutionary synthesis models}

The second conclusion is supported by evolutionary synthesis modeling of the 
spectrophotometric data of regions III to VI.
To estimate quantitalively the age of each region, we calculate a grid 
of  spectral energy distributions (SED) 
for stellar populations with ages varying between 10 Myr and 20 Gyr and heavy
element abundance $Z_\odot$/20,  using isochrones from
Bertelli et al. (1994) and the compilation of stellar atmosphere models from
Lejeune et al. (1998). A Salpeter IMF with slope --2.35, an upper 
mass limit of 120 $M_\odot$ and a lower
mass limit of 0.6 $M_\odot$ were adopted. For stellar populations with
 age less than 10 Myr we use theoretical spectral energy distributions by
Schaerer \& Vacca (1998) for a heavy element abundance $Z_\odot$/20
and a Salpeter IMF. The stellar emission in SBS 1415+437
is contaminated by emission of ionized gas from supergiant H II regions.
Therefore, to study the stellar composition in the BCD,  it is necessary
to produce a synthetic SED which includes both stellar and ionized gaseous
emission. It is clear from Figure 10, the best fit to the 
spectrum corrected for extinction (lower panel)
 gives an age between 30 and 100 Myr for region V (the 
models are labelled by the logarithm of the age in years). A similar analysis for 
the other regions give the same answer: none contains stellar populations older than
100 Myr. Thus SBS 1415+437, just like SBS 0335--052,  is also a young galaxy.

\section{Dust in an extremely metal-deficient environment}

\def\baselinestretch{0.5}
\begin{figure*}[t]
 \vbox{
	\includegraphics{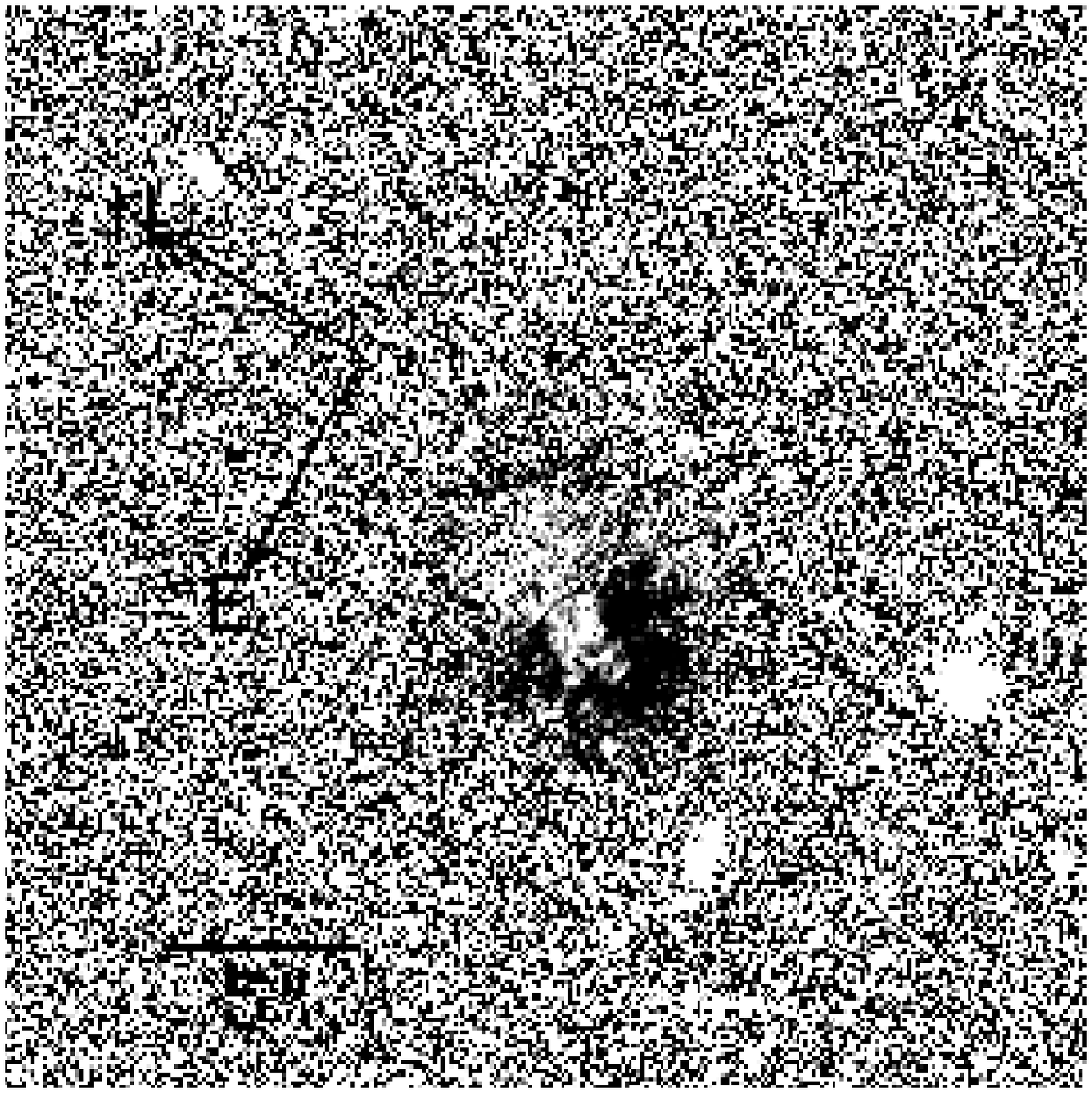}}
	\includegraphics{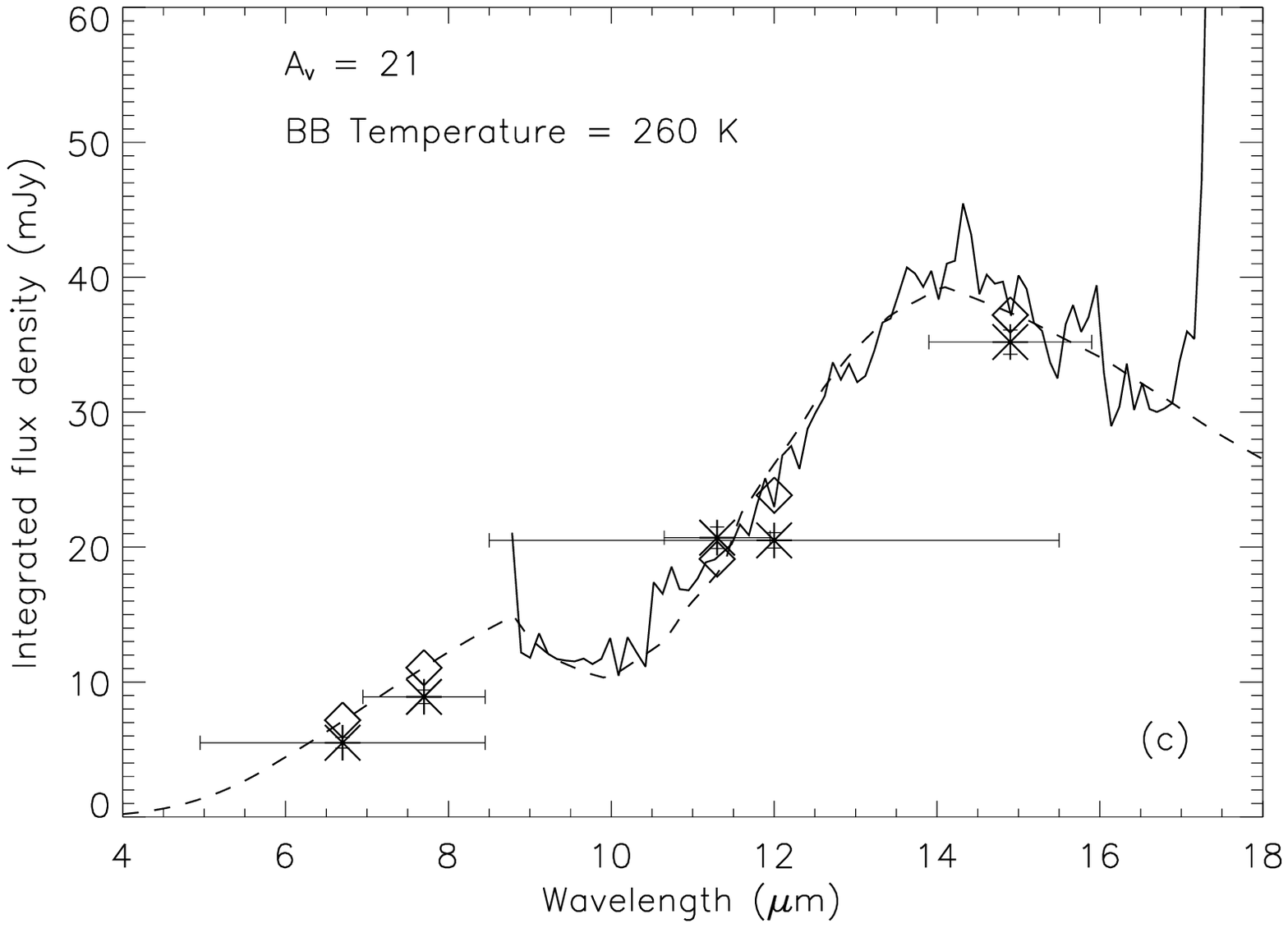} 
\vspace{8.5cm}
{\label{}{\bf Fig.11 (left):}
$(V-I)$ color map of SBS 0335--052. Blue is dark, red is light. The dust patches on 
top of the super-star clusters are clearly seen.

{\bf Fig.12 (right):} The best-fit model (thick continuous line) for the mid-infrared spectral
energy distribution of SBS 0335--052
obtained by ISO (Thuan et al. 1999).  It is composed of a 260\,K blackbody
spectrum modified by an emissivity law proportional to $\nu^{1.5}$ and
extinguished by a screen of dust of optical thickness $A_{V}$ = 21 mag,
with an extinction curve similar to that observed 
toward the Galactic center  (Lutz et al. 1996).}  
\end{figure*}
\def\baselinestretch{1.0}

SBS 0335--052, although it is extremely metal-deficient, clearly contains dust. 
It is seen as 
white patches in the $V-I$ color map of the BCD (Figure 11), and is spatially 
mixed with the SSCs.  
Thuan et al. (1999) have obtained ISO mid-infrared observations of the BCD.
  With 
$L_{12{\mu}m}/L_{B}$ of 2.15, the galaxy is unexpectedly bright in the 
mid-infrared for such a low-metallicity object. The mid-infrared 
spectrum (Figure 12) is very unusual when compared to spectra of other
 star-forming
galaxies: 1) there is no emission from the so-called Unidentified 
Infrared Bands, which we interpret 
as an effect of destruction of carbon-based dust by the very high UV 
energy density in SBS 0335--052 ; and 2) there are 
no evident fine structure ionic lines, even though neon and sulfur lines 
are usually quite bright in starburst galaxies. The absence of these lines can 
be explained by a very strong continuum which decreases the equivalent 
widths of the lines, making them difficult to detect. 
  The spectral energy distribution from 5 to 
17 microns can be fitted with a grey-body spectrum modified by extinction.
Silicate grains can account for the unusual shape of the MIR spectrum. The 
required extinction law is similar to that observed toward the 
Galactic Center and the optical depth is $A_{V}\,\sim$\,19-21 mag. Such a 
large optical depth implies that a large fraction (as much as 75\%) of the 
current star-formation activity in SBS 0335--052 is 
hidden by silicate dust whose mass is in the range 3$\times$10$^3$ -- 
5$\times$10$^5$ $M_\odot$. 

Thus nearly primordial environments can contain a significant amount of dust.
If the hidden star formation in SBS 0335--052 is typical of young galaxies at 
high redshifts, then the cosmic star formation rate derived from UV/optical 
fluxes would be underestimated.
This is in fact the result found by Flores et al. (1999) who carried out  
deep 15 micron ISO surveys of distant galaxies with a 
median redshift of $\sim$ 0.76. Those authors found that the 
cosmic SFR derived from FIR luminosities, from 43 to 123 microns, 
assuming that the FIR luminosity comes mostly from dust heating by young stars, 
 is 2-3 times higher  than the SFR 
estimated previously from optical/UV fluxes (Madau et al. 1996).

\section{The escape of Ly$\alpha$ photons}

   The study of nearby young galaxies can also shed light on a long standing
problem concerning the Lyman-alpha emission of primeval galaxies. In any
galaxy formation scenario, young galaxies are predicted to show strong
Ly$\alpha$ emission (rest frame equivalent width of $\sim$ 100\AA) associated
with the formation of a large number of massive ionizing stars (e.g. Charlot \&
Fall 1993). Yet, despite intensive searches, the predicted population of 
Ly$\alpha$ primeval galaxies remained elusive (Pritchet 1994). The absence of
Ly$\alpha$ emission in high-redshift galaxies is reminiscent of the behavior
of Ly$\alpha$ emission in nearby 
starburst and BCD galaxies, which is either absent
or greatly diminished. 
The Ly$\alpha$/H$\beta$ line intensity ratio in those
galaxies with detected Ly$\alpha$ emission does not exceed 10, significantly
lower than the theoretical recombination ratio of 33. Some galaxies show strong
Ly$\alpha$ absorption rather than emission. 
The favored explanation for such a reduction in Ly$\alpha$
emission from recombination values is redistribution of Ly$\alpha$ 
photons by multiple scattering in the H I envelope or 
absorption of these  
Ly$\alpha$ photons by dust in the star-forming region.
The latter mechanism would imply increasing Ly$\alpha$/H$\beta$ line intensity 
ratios with decreasing metallicities, since presumably low-metallicity
objects contain less dust, and hence suffer less destruction of Ly$\alpha$
photons (Terlevich et al. 1993).
{ \sl HST} observations of the two most metal-deficient
BCDs known, I Zw 18 (Kunth et al. 1994) and SBS 0335--052 (Thuan et al. 
1997, section 4) , show Ly$\alpha$ not in emission but absorption
(Figure 7), which goes against a
Ly$\alpha$ strength-metallicity anticorrelation.
Lequeux et al. (1995), in their study of the BCD Haro 2 ($Z_\odot$/3) which 
shows Ly$\alpha$ in emission, have
argued that Ly$\alpha$ photons can escape when the neutral material where
the absorption occurs is outflowing with a velocity of $\leq$ 
200 km s$^{-1}$ with respect to the star-forming region. 
The Ly$\alpha$ emission is redshifted with
respect to both the H II region and the expanding absorbing shell, the motion
of which is probably powered by stellar winds and supernovae. 
This explanation does not apply to the BCD 
T1214--277 ($Z_\odot$/23) which is the
lowest metallicity galaxy known with detected Ly$\alpha$ emission (Thuan
\& Izotov 1997). 
Contrary to Haro 2,
 Ly$\alpha$ emission is not redshifted with respect to the H II gas velocity.
Thus the escape of Ly$\alpha$ photons in T1214--277 is not a consequence of the
motions of the neutral H I envelope. 

As for  SBS 0335--052, dust extinction may play some role as it is
directly seen (Figure 11). While there is
evidence for fast gas motions in SBS 0335--052 with velocities up to 
$\sim$ 1500 km s$^{-1}$, the H I gaseous envelope appears to be static with
respect to the H II region, as the 21 cm and emission-line velocities are in
good agreement. Thus, with its extremely large H I column density, the
redistribution of Ly$\alpha$ photons in SBS 0335--052 by multiple scattering
over the large volume of the H I cloud probably plays also an important role
in diminishing the intensity of the Ly$\alpha$ line. The orientation of the 
HI cloud may also play a role. In the case of SBS 0335--052 , the HI envelope
 is reasonably flattened (Pustilnik et al. 1999) suggesting it is seen nearly 
edge-on. In that case, Ly$\alpha$
photons escape more easily along directions perpendicular to the line of sight 
than along it. Moreover, as discussed by Giavalisco et al. (1996), the escape of
 Ly$\alpha$
photons may be controlled not only by the geometry but also by the
porosity of the neutral gas. 

In summary, there is no unique mechanism which controls the appearance
of Ly$\alpha$ emission in nearby young dwarf galaxies. Dust extinction may
play a role, but the velocity structure of the H I gas, the orientation of
the H I cloud and its porosity may also be important factors.
The fact that some  BCDs do not show Ly$\alpha$ in emission implies that 
Ly$\alpha$ searches for high-redshift galaxies will always be incomplete.

\acknowledgements{We thank the partial financial support of NSF grant 
AST-9616863 and NASA grant JPL961535, and Polichronis Papaderos, Simon 
Pustilnik and Marc Sauvage for useful discussions.}


%
\vfill
\end{document}